\documentclass[aps,prd,preprint]{revtex4}
\usepackage{epsfig}
\begin{document}

\newcommand{\E}{{\cal E}}
\newcommand{\F}{{\cal F}}
\newcommand{\A}{{\cal A}}
\newcommand{\alr}{\alpha r}
\newcommand{\asr}{\alpha^{2}r^{2}}
\newcommand{\dspst}{\displaystyle}

\title{Four-dimensional anti--de Sitter toroidal black holes from a
three-dimensional  perspective: full complexity}
\author{Vilson T. Zanchin\footnote{email: zanchin@ccne.ufsm.br}}
\affiliation { Departamento de F\'{\i}sica, Universidade Federal de
Santa Maria, 97119-900 Santa Maria, RS, Brazil}
\author{Antares Kleber\footnote{anta@on.br}}
\affiliation {Observat\'orio Nacional -- MCT,
Rua General Jos\'e Cristino 77, 20921 Rio de Janeiro, RJ, Brazil,}
\author{Jos\'e P. S. Lemos\footnote{lemos@kelvin.ist.utl.pt}}
\affiliation{Centro Multidisciplinar de Astrof\'{\i}sica --
CENTRA, Departamento de F\'{\i}sica,
Instituto Superior T\'ecnico, Av. Rovisco Pais 1, 1096 Lisboa, Portugal\\
and  Observat\'orio Nacional -- MCT,
Rua General Jos\'e Cristino 77, 20921 Rio de Janeiro, RJ, Brazil}

\vskip 1cm

\begin{abstract}

\noindent

The dimensional reduction of black hole solutions in four-dimensional
(4D) general relativity is performed and new 3D black hole solutions
are obtained.  Considering a 4D spacetime with one spacelike Killing
vector, it is possible to split the Einstein-Hilbert-Maxwell action
with a cosmological term in terms of 3D quantities.  Definitions of
quasilocal mass and charges in 3D spacetimes are reviewed.  The
analysis is then particularized to the toroidal charged rotating
anti-de Sitter black hole.  The reinterpretation of the fields and
charges in terms of a three-dimensional point of view is given in each
case, and the causal structure analyzed.

\medskip

\end{abstract}

\maketitle



\section {Introduction}

The work on three-dimensional (3D) gravity theories has seen a great
impulse after the discovery that 3D general relativity possesses a
black hole solution, the Ba\~nhados-Teitelboim-Zanelli (BTZ) black 
hole \cite{btz,bhtz}.  Before the
appearance of this black hole solution there were, however, important
works in 3D general relativity which studied the
properties of point particles in 3D geometries \cite{deser1,deser2}
as well as solutions with matter \cite{giddings}, and 
showed that it provides a testbed for
4D and higher-D theories \cite{achucarro,witten}. In addition there
were works in 3D string theory with its associated black strings
\cite{horowitz1,horowitz2}.  However, because of the lack of black holes
there was no possibility of discussing important issues such as the
entropy of the gravitational field, its degrees of freedom, and
Hawking evaporation.
The work by Ba\~nados, Teitelboim, Zanelli, and Henneaux
\cite{btz,bhtz} brought then
3D general relativity into the level of complexity of 4D general
relativity. This black hole is a solution of the Einstein-Hilbert
action including a negative cosmological constant term $\Lambda$.  One
can also show that the BTZ black hole can be constructed by
identifying certain points of the 3D anti--de Sitter (AdS) spacetime
\cite{bhtz}. Since the AdS spacetime is a simple manifold one can
study many properties of the  BTZ black hole through known results in
AdS spaces, upon making further appropriate global identifications
(see \cite{carlip} for a review).

After the BTZ solution a whole set of new solutions in 3D followed
from a number of different dilaton-gauge vector theories coupled to
gravity. For instance, upon reducing 4D Einstein-Maxwell theory with
$\Lambda$ and with one spatial Killing vector it was shown in
\cite{lemos1,lz1} that it gives rise to a 3D Brans-Dicke-Maxwell
theory with its own black hole, which when reinterpreted back in 4D
is a black hole with a toroidal horizon. One can then naturally extend
the whole set to Brans-Dicke theories \cite{sa,dias1}. Other solutions
with different couplings have also been found \cite{chan,martinez} (see
\cite{dias1} for a more complete list).

One important ingredient in these solutions is the presence of a
negative cosmological constant, $\Lambda<0$. The interest in these
solutions appeared after it was shown that gauged supergravity
requires for its ground state a $\Lambda<0$ term, such that spacetime
is AdS or asymptotically AdS. Many of these black
holes, such as the BTZ black hole, only exist in theories with a
negative $\Lambda$ \cite{ida}, which in turn motivated their
further study.
A renewal of interest in these solutions  came after the
AdS-conformal field theory (AdS-CFT) conjecture \cite{maldacena}. This
conjecture states the equivalence between string theory on an AdS
background and a corresponding CFT defined on the boundary of AdS
spacetime, i.e., between AdS$_{ n}$ and a CFT$_{n-1}$.  The $n=3$ case
(i.e., 3D), mainly through the BTZ black hole, plays an important role in
the verification of the conjecture, since many higher-D extreme black
holes of string theory have a near-horizon geometry containing the BTZ
black hole.  Then the conjecture says that if one has, e.g., a
10-dimensional type IIB supergravity, compactified into a ${\rm BTZ}
\times S^3 \times T^4$ spacetime, the BTZ in the bulk corresponds to a
thermal state in the boundary CFT
\cite{maldacenastrominger}. It is also possible to
embed the 4D toroidal black holes \cite{lemos1,lz1}
in a higher-D string theory such that they can be interpreted
as the near horizon structures of an $M2$ brane rotating in extra
dimensions \cite{duff}.

Now, in order to find solutions in a given dimension, $n$ say, one
usually starts from the action of gravity theory with the generic
dilaton and gauge vector couplings in that dimension, derives the
corresponding equations of motion, tries an ansatz for the
solution and then finally from the differential equations finds
the black hole solutions compatible with the ansatz.  This is the
case for the Schwarzschild solution for instance, and for many of
the 3D solutions quoted above. Another way of finding solutions
arises if the theory possesses dualities, i.e., symmetries that
convert one solution into another in a nontrivial way
\cite{horowitzreview92}. Yet another way, which can be seen as a
special case of duality, is through dimensional reduction, where
one can reduce a theory by several dimensions. The simplest case
is to reduce by one dimension, i.e., one starts with a ($n+1$)D
Lagrangian theory and through a suitable procedure reduces it
along a symmetry direction into a new $n$D Lagrangian theory.
There are a number of inequivalent procedures to perform a
dimensional reduction, two of those are the dimensional reduction
through a Kaluza-Klein ansatz (or classical Kaluza-Klein
reduction) \cite{duff2}, and the Lagrangian dimensional reduction
\cite{cremmer}. When one is reducing through one symmetric compact
direction (a circle), which will be the cases studied here, both
procedures are equivalent \cite{duff2,breiten}.  In the reduction
process scalars and gauge vector fields appear naturally. The
symmetry direction of the solution in the ($n+1$)D theory defines
a Killing vector and a Killing direction, which in general can be
compact or non-compact. In turn, in the non-compact case the
reduction process can be important to the original theory in
($n+1$)D.  For example, for a black hole solution in the lower
$n$D theory one can give precise definitions of charges (mass,
angular momentum, electromagnetic, dilaton and axion charges),
which in the ($n+1$)D theory are then converted into charges per
unit length of the corresponding black string
\cite{lemos1,abicaks}.  In the past two decades, due to the extra
dimensions required by supergravity and string theories, the
techniques of compactification and dimensional reduction have
became powerful tools to build and analyze black hole solutions in
lower dimensions (see, e.g., 
\cite{horowitzreview92,cremmer,duff2,breiten,abicaks,KSS}).

In this paper we follow the classical Kaluza-Klein procedure, here
equivalent to the Lagrangian dimensional reduction procedure, to find
new solutions in 3D from solutions in 4D.  An extensive study of 4D
solutions of gravity coupled to sigma model theories and their
corresponding Kaluza-Klein 3D counterparts has been performed
\cite{breiten}, in which, in the last section of the paper on ``Open
Problems'' the authors state that the inclusion of a cosmological
constant is important. It is our aim to apply the dimensional
reduction technique to construct 3D black holes from the 4D toroidal 
AdS black holes \cite{lemos1,lz1}.  Sect. II is dedicated to a review of the
dimensional reduction method particularized to the case of reduction
from 4D to 3D. The definition of the charges for all fields, new and
old, appearing in 3D is also given.  Then, in Sect. III,  
dimensional reduction, through
the Killing azimuthal direction $\partial/\partial\varphi$, of
rotating charged black holes with toroidal topology is considered. 
The produced 3D black holes display an isotropic horizon
(i.e., circularly symmetric), and the new charges are neatly found.
In Sect. IV we conclude.

\section{Dimensional Reduction}\label{sectdefs}

\subsection{The action, the Lagrangian and the fields}

In this subsection we discuss the connection between the
Einstein-Maxwell-AdS equations in a four-dimensional (4D) spacetime with a
Killing vector and the equations in the corresponding 3D spacetime.
We assume that the 4D manifold ${\cal M}_4$ can be decomposed as
${\cal M}_4={\cal M}\times {\cal S}^1$ or ${\cal M}_4={\cal M}\times R$, with
$\cal M$, ${\cal S}^1$, and $R$ being the 3D manifold, the circle, and the
real line, respectively.

The action $\hat S$ in 4D spacetimes is assumed to be the usual
Einstein-Hilbert-Maxwell action with cosmological term $\hat
\Lambda$ and electromagnetic field $\hat {\bf F} = {\rm d} \hat {\bf
A}$ (where $\hat {\bf A}$ is the gauge field), given by  (we use geometric
units where $G=1$, $c=1$), 
\begin{equation} \hat S = 
\int d^4x \, {\hat {\cal L}}=
\frac{1}{16\pi}\int{d^{4}x\sqrt{-\hat g} \left(\hat R-2 \hat \Lambda -\hat
F^2\right)}\, ,                         \label{lagrN1}
\end{equation}
where ${\hat {\cal L}}$ is the Lagrangian density, or Lagrangian.  
The convention adopted here is that quantities wearing hats are defined in
4D and quantities without hats belong to 3D manifolds.  We now proceed with
the reduction of the action 
(\ref{lagrN1}) to 3D. In order to do that consider then a 4D spacetime
metric admitting one spacelike Killing vector, $\dspst{\partial_\varphi}$,
where $\varphi$ can be a compact or a non-compact direction. In such a case
the 4D metric may be decomposed into the form
\begin{equation}
d{\hat s}^2=e^{2 \beta_0 \phi}ds^2
+e^{2\beta_1\phi}\left(d\varphi+\A_{i}dx^{i}\right)^2 \, ,
\label{metricN1}
\end{equation}
where $ds^2$ is the 3D metric, $\phi$,
$\A_{i}$ ($i= 0,1, 2$) and all the other metric coefficients are
functions independent of $\varphi$, and $\beta_0$, $\beta_1$ are numbers.
To dimensionally reduce the electromagnetic gauge field we do
\begin{equation}
\hat {\bf A} = {\bf A} + A_\varphi {\rm\bf d}\varphi\,,
\label{reducedA}
\end{equation}
where for compact $\varphi$ the gauge group is $U(1)$ and for
non-compact $\varphi$ it is $R$.  In the last equation $\bf A$ is a
$1$-form while $A_\varphi$ is a $0$-form. In terms of a coordinate
basis in the 3D manifold this means that $\bf A$ and $A_\varphi$ correspond
to a vector
field $A_i$ and to a scalar field $A_\varphi=\Psi$, say,
respectively. From these fields we can define the 3D Maxwell field
\begin{equation}
\bf F = {\rm {\bf d}}\bf A\,.
\label{F}
\end{equation}
It is convenient to define a new 2-form $\bf E$ as
\begin{equation}
{\bf E} \equiv {\bf F} -{\bf d} \Psi \wedge \A \,,\label{GN}
\end{equation}
 where $\A$ is the Kaluza-Klein
gauge field 1-form appearing in the 4D metric (\ref{metricN1}).

The Kaluza-Klein dimensional reduction procedure then gives
\begin{eqnarray}
S& = &\frac{L_3}{16\pi }\int {d^{3}x\sqrt{- g}
e^{(\beta_0+\beta_1)\phi}
\left[R-2 \Lambda e^{2\beta_0\phi}+2\beta_0(\beta_0+2\beta_1)
\left(\nabla\phi\right)^2 \right.}\nonumber\\
& &\left. -{}e^{-2\beta_0\phi}E^2
-{1\over 4} e^{2(\beta_1-\beta_0)\phi}\F^2
-{2}e^{-2\beta_1\phi} ({\bf \nabla} \Psi)^2\right]\, ,  \label{lagrN}
\end{eqnarray}
where we have defined
\begin{equation}
\F = {\bf d} \A\,.
\label{calF}
\end{equation}
The 3D cosmological constant is defined
as $\Lambda=\hat \Lambda$.
$L_3$ is the result of integration
along the $\varphi$ direction.  Assuming the spacetime is
compact along the generic spacelike dimension parametrized by $\varphi$,
then $L_3$ is given by the range of $\varphi$
and will be dimensionless, while for noncompact $\varphi$, $L_3$
carries physical units of length.
{From} the 3D point of view, $L_3$ can be thought of as the size of
the extra dimension. 
The explicit form of  $E^2$
in terms of the 3D fundamental fields $F_{ij}$, $\A_i$ and $\Psi$ is
\begin{equation}\label{Eexpl}
E^2 \equiv E_{ij}E^{ij}= F^2+ 4F^{ij}\A_i\,\nabla_j\Psi
+ 2\left[({\bf\nabla} \Psi)^2\A^2 -
(\A^i\,{\nabla_i}\Psi)^2\right]\, .
\end{equation}
The reduced action shows two different gauge fields. The first
one, $A_i$, is the 3D counterpart of the 4D electromagnetic gauge
field. The second, $\A_i$, is the Kaluza-Klein gauge field. There are
also two scalar fields. The dilaton $\phi$, and another scalar
field $\Psi$, which
is the projection of the 1-form gauge field $\hat{\bf A}$ onto the
Killing direction $\varphi$.  The scalar field $\Psi$ couples to the metric
differently from a true scalar field. This can be seen, for instance,
by  comparing the
last term in the action (\ref{lagrN}) to the kinetic term for the
dilaton field $(\nabla \phi)^2$.

The equations of motion which follow from the action (\ref{lagrN}) for the
graviton $\bf g$, the gauge fields $\bf A$ and $\cal A$, the dilaton $\phi$,
and the scalar $\Psi$ are, respectively:
\begin{eqnarray}
&  & G_{ij}= - \Lambda e^{2\beta_0\phi}g_{ij}+ (\beta_0+\beta_1)\left[
\nabla_i \nabla_j\phi  -(\beta_0+\beta_1)\nabla_i\phi\nabla_j\phi- g_{ij}
\nabla^2\phi\right]\nonumber\\  & &\hskip 1.2cm
+\beta_1^2  \left[2\nabla_i\phi\nabla_j\phi -
g_{ij}\left(\nabla\phi\right)^2\right]
 -{2}e^{-2\beta_0\phi}\left[ E_{ik} E^k_j +{1\over
4}g_{ij}E^2\right]  \nonumber \\ & & \hskip 1.2cm
-{1\over 2}e^{2(\beta_1-\beta_0)\phi}
\left[\F_{ik}\F^k_j+{1\over 4}g_{ij}\F^2\right]
+{2}e^{-2\beta_1\phi}\left[
\nabla_i\Psi\nabla_j\Psi  -{1\over 2}g_{ij}(\nabla\Psi)^2\right]
 \, , \label{eisteineq}\\ & &
\nabla_j \left[e^{(\beta_1-\beta_0)\phi}\left(F^{ij}- \A^i\nabla^j\Psi
+\A^j\,\nabla^i\Psi\right)\right] =0 \label{gaugefieldeom} \, ,\\
& & \nabla_j\left[e^{(3\beta_1 -\beta_0)\phi}\F^{ij}\right]=
4e^{(\beta_1-\beta_0)\phi}\left[-F^{ij}\nabla_j\Psi + \A^i
\left(\nabla\Psi\right)^2- \left(A^j \nabla_j\Psi \right)\nabla^i\Psi\right] \,
,\label{kkgaugefieldeom} \\
 & &\nabla_i\left[\beta_1^2 e^{(\beta_0+\beta_1)\phi}\nabla^i\phi\right]  =
-\Lambda\,\beta_1e^{(\beta_1+3\beta_0)\phi}
+{1\over 4}\beta_1{e^{(3\beta_1-\beta_0)\phi}}\F^2 \nonumber \\  & &
\hskip 1.3cm  +  {1\over 2}\beta_1 e^{(\beta_1-\beta_0)\phi}E^2
 - \beta_1{e^{(\beta_0-\beta_1)\phi}}\left(\nabla\Psi\right)^2
\, , \label{phieom} \\
& & \nabla_i\left[e^{(\beta_0-\beta_1)\phi}\nabla^i\Psi
+ e^{(\beta_1-\beta_0)\phi} \left( F^{ij}\A_j +\A^2\nabla^i\Psi
 -\A^i\, \A^j\nabla_j\Psi\right)\right] = 0\, ,\label{psieom}
\end{eqnarray}
where $G_{ij}$ is the Einstein tensor.

Now, we are free to choose $\beta_0$ and $\beta_1$, i.e., we are free
to choose the frame in which to work, with the different frames being
related by conformal transformations.  There are three frames that
stand out:

(i) the good frame, i.e., the one that preserves most of
the structure of the 4D spacetimes, is given
by $\beta_0=0$ and $\beta_1$ free, which we can normalize
to $\beta_1=-2$, yielding the following action $S_{\rm g}$
(for these three particular cases, we do not display the 
equations of motion, only the action since
it is much more condensed)
\begin{eqnarray}
S_{\rm g}& = &\frac{L_3}{16\pi}\int {d^{3}x\sqrt{- g} e^{-2\phi}
\left[R-2 \Lambda -E^2 -{1\over 4} e^{-4\phi}\F^2
-{2}e^{4\phi} ({\bf \nabla} \Psi)^2\right]}\, ;  \label{lagrNgood}
\end{eqnarray}
(ii) the Einstein frame, the one that preserves the
Einstein form of the action, is given by
choosing $\beta_0+\beta_1=0$, and
$\beta_0=1/2$, so that the kinetic term of the dilaton is
$2\beta_0(\beta_0+2\beta_1)=-1/2$ (see
e.g. \cite{KSS} for dimensional reduction in the Einstein frame),
yielding the  action $S_{\rm E}$
\begin{eqnarray}
S_{\rm E}& = &\frac{L_3}{16\pi}\int {d^{3}x\sqrt{- g}
\left[R-2 \Lambda e^{\phi}-{1\over 2} \left(\nabla\phi\right)^2
 -e^{-\phi}E^2 -{1\over 4} e^{-2\phi}\F^2
 -{2}e^{\phi} ({\bf \nabla}
\Psi)^2\right]}\, ; \label{lagrNeinstein} \end{eqnarray}
(iii) the string frame, where one chooses $\beta_0+\beta_1=-2$ and
$\beta_0=-2\pm\sqrt2$,  fixing the kinetic term of
the dilaton to $2\beta_0(\beta_0+2\beta_1)=4$,
yielding the following action $S_{\rm s}$
\begin{eqnarray}
S_{\rm s}& = &\frac{L_3}{16\pi}\int {d^{3}x\sqrt{- g} e^{-2\phi}
\left[R-2 \Lambda e^{-2(2 \mp\sqrt{2})\phi}+4
\left(\nabla\phi\right)^2 \right.}\nonumber\\
& &\left. -e^{2(2 \mp\sqrt{2})\phi}E^2
 -{1\over 4} e^{4(1\mp\sqrt{2})\phi}\F^2
-{2}e^{\pm 2\sqrt{2}\phi} ({\bf \nabla} \Psi)^2\right]\, .
\label{lagrNstring} \end{eqnarray}
It is well known that the different frames, related by conformal
transformations, are physically inequivalent, e.g., one frame can give
spacetime singularities where the other does not (see, e.g.,
\cite{chan-mann,marolf}).  We will work mainly with the good frame, and
we will comment later on the other frames.

As it will be seen later on, the 3D solutions are obtained from the
4D metric and the other 4D fields  by direct inspection of the metric
and correct truncation of extra fields. This task is more easily
accomplished by working in the good frame.  Once we have the 3D metric in
the good frame ($\beta_0=0$), the metric in any other frame can be
obtained by the appropriate conformal transformation. In order to
build such a transformation, let us define $ds^2_{\rm g}$ and 
$ds^2_{\rm o}$
as the given metric written in the good and other frames,
respectively.  In the good frame, the parameters $\beta_0$ and $\beta_1$
assume respectively the values $(\beta_0)_{\rm g}$ and $(\beta_1)_{\rm g}$ 
[as we
have mentioned, we chose $(\beta_0)_{\rm g}=0$ and $(\beta_1)_{\rm g}=-2$]. 
Let us also denote the values assumed by the parameters $\beta_0$ and
$\beta_1$ in the other frame, respectively, by $(\beta_0)_{\rm
o}$ and $(\beta_1)_{\rm o}$. The two frames are then related by
\begin{equation}
ds^2_{\rm o}= \left(e^{-2\beta_1\phi}\right)
_{\!{\rm g}}^{\left(\beta_0/\beta_1\right)_{\!\rm o}} ds^2_{\rm g}\,,
\label{conftransf}
\end{equation}
where $ \left(e^{2 \beta_1\phi}\right)_{\rm g}$ is the dilaton field 
in the good
frame. For instance, 
the relation between good and Einstein frames is
$ds^2_{_{\rm E}} = \left(e^{2 \beta_1\phi}\right)_{\!{\rm g}}ds^2_{\rm g}$.

\subsection{The global charges}

Now we study how to define mass, angular momentum  and charges in the 3D 
spacetime  by using the formalism of Brown and York 
\cite{by1,brown-mann,creighton-mann} modified to include a dilaton and 
other fields.

\subsubsection{The conventions}
We assume that the 3D spacetime  $\cal M$ is topologically the
product of a spacelike surface $D_2$ and a real line time interval $I$,
${\cal M} =D_2\times I$.  $D_2$ has the topology of a disk. Its boundary
$\partial D_2$ has the topology of a circle and is denoted by ${\cal S}_1$.
The boundary of $\cal M$, $\partial \cal M$, consists of
two  spacelike  surfaces $t=t_{1}$ and $t=t_{2}$, and a timelike
surface ${\cal S}_1\times I$ joining them. Let $t^i$ be a timelike
unit  vector ($t_it^i=-1$)
normal to a spacelike surface $D_2$ (that foliates $\cal M$), and $n^i$
be the outward unit vector normal to the boundary $\partial\cal M$
($ n_in^i=1$). Let us denote the
spacetime  metric on $\cal M$ by  $g_{ij}$ ($i,j=0,1,2$).
 Hence $h_{ij}= g_{ij} + t_it_j$ is the induced metric on $D_2$
and $\sigma_{ij}= g_{ij}+t_it_j-n_i n_j$ is the  induced metric on
${\cal S}_1$.  $h_{ij}$ can be viewed also as a tensor $h_{mn}$ ($m,n=1,2$)
on $D_2$, and $\sigma_{ij}$ can be viewed as a scalar (a tensor of rank
zero)  $\sigma_{ab}\equiv\sigma$ ($a,b=2$) on the one-dimensional boundary
${\cal S}_1$. Since ${\cal S}_1$
is a one-dimensional space, the induced metric $\sigma_{ab}$ has only
one independent component.
The induced metric on the spacetime boundary $\partial \cal  M$ is
$\gamma_{ij}= g_{ij}-n_i n_j = \sigma_{ij}-t_i t_j$.
We also assume that the spacetime admits the
two Killing vectors needed  in order to define mass and angular
momentum: a timelike Killing vector
$\eta_{t}^{i}=({\partial}/{\partial t})^{i}$ and a spacelike
(axial) Killing vector
$\eta_{\theta}^{i}=(\partial_\theta)^{i}$.

\subsubsection{Mass}
The next step is to adapt the Brown and York procedure to take into
account the dilaton field \cite{creighton-mann}.  By doing this, we
arrive at the following
definition of mass $M$ on a 3D spacetime admitting a timelike Killing
vector $\eta_t$ 
\begin{equation} M_{3D} = {L_3\over
8\pi}\int_{{\cal S}_1}e^{2(\beta_0+\beta_1)\phi}
\delta\left({k^\phi}\right)t_{i} \eta_t^{i}\,d{\cal S}\, ,
\label{massdef} \end{equation}
where $t^{i}$ is the timelike future pointing normal to $D_2$,
$d{\cal S} =\sqrt{\sigma}\,d\xi$ with $\xi$ being a coordinate on
${\cal S}_1$, and $\sigma$ being the determinant of the induced
metric on ${\cal S}_1$ (since ${\cal S}_1$ is a one-dimensional
space, the determinant $\sigma$ of the induced metric
$\sigma_{ab}$ coincides with the metric itself). ${k^\phi}$ is the
trace of the extrinsic curvature of ${\cal S}_1$ as embedded on
$D_2$, modified by the presence of the dilaton.  To define
${k^\phi}$ explicitly we consider the particular case when the
two-metric on $D_2$ can be split as
\begin{equation}
ds_{D_2}^2 = h_{mn}dx^m\,dx^n = f^{2}dr^2 +
R^2\left(d\xi+V\,dr\right)^2\,, \label{D2metric}
\end{equation}
where $m,n=1,2$, $x^1 =r$, $x^2=\xi$, and $\xi$ parametrizes ${\cal S}_1$.
Functions  $f$, $R$ and $V$ depend on  all  coordinates.
$k^{\phi}$ may then be written as
\begin{equation} {k^{\phi}}= -{e^{-(\beta_0+\beta_1)\phi}\over 2}{1\over f}
\left({2\over R} {\partial R\over\partial r} + 2\left(\beta_o+\beta_1\right)
{\partial \phi \over\partial r}
-\nabla_\xi V \right)\, ,
\label{kappadef}
\end{equation}
where $\nabla_\xi$ is the covariant derivative on ${\cal S}_1$.
Recall that the energy surface density on ${\cal S}_1$, $\epsilon$, is
given by
\begin{equation}
\epsilon =\frac{k^\phi}{8\pi}\, . \label{energydensity}
\end{equation}
An explicit definition of $\epsilon$ is given for the particular case
studied in Sect. \ref{secttoroidal} below. In Eq. (\ref{massdef}), the
symbol $\delta$ indicates the difference between the extrinsic
curvature $k^\phi$ on the spacetime $\cal M$ obtained from the
full action $S$, say, and the corresponding quantity
$\left(k^\phi\right)_{\!o}$ obtained from a reference spacetime
${\cal M}_o$, solution of a reference action $S_o$.  Namely,
$\delta({k^\phi})= {k^\phi}- ({k^\phi})_o$. In this paper we are
interested in black holes in asymptotically AdS spacetimes. Hence,
the action $S$ refers to a specific black hole solution as, e.g.,
the toroidal rotating charged-AdS spacetime, and $S_o$ refers to
the (asymptotic) AdS spacetime, when no black hole is present. The
right hand side of Eq. (\ref{massdef}) is  the quasilocal mass as
defined in Brown-York formalism and in general depends on the
choice of the boundary ${\cal S}_1$. In our definition we assume
that ${\cal S}_1$ represents the infinite boundary of the
two-space $D_2$, and the integration over ${\cal S}_1$ then gives
the global mass associated with the considered black hole solution.

\subsubsection{Angular momentum}
Similarly to the  mass, the definition of  angular momentum $J$
for a 3D spacetime admitting a spacelike Killing vector
$\partial_\theta=\eta_\theta$ can also be modified to include the dilaton.
The definition of the angular momentum is then 
 \begin{equation}
J_{3D}= {L_3}\int_{{\cal S}_1}
e^{2(\beta_0+\beta_1)\phi}\delta\left({j^{\phi}}_{i}
\right)\eta_\theta^{i} \,d{\cal S}\, ,
\label{momentumdef}
\end{equation}
where ${j^{\phi}}_{i}$ is the momentum
surface density on ${\cal S}_1$, modified by the presence of the dilaton.
We also have  $\delta ({j^{\phi}}_i)\equiv (j^{\phi}_i)-(j^{\phi}_i)_o$,
where $(j^{\phi}_i)_o$ is the angular-momentum density at
the boundary ${\cal S}_1$ of the background (or reference) spacetime, and
${j^{\phi}}_i$ is the full angular momentum density of the considered
spacetime.  We do not give here an explicit definition for ${j^{\phi}}_i$,
since it will not be needed in the applications considered in the present
work.

Since we are interested in the global conserved quantities of
black holes, in Eq. (\ref{momentumdef}), as in  Eq.
(\ref{massdef}) and in the charges defined below, 
the integral over the boundary of $D_2$, ${\cal
S}_1$, is in fact taken at the infinity of $D_2$, ${\cal S}_1
\longrightarrow \infty$. In such a limit, the mass, angular
momentum, and charges defined according to the Brown and York
formalism coincide with the ADM mass, angular momentum  and
charges.

Now we turn our attention to the definition of other charges in 3D
spacetimes.

\subsubsection{Electric and magnetic charges}

{\it (a) \ Electric charges of the gauge fields $A_i$ and
$\A_i$}.
The two gauge fields, $A_i$ and $\A_i$, have different electric
charges $Q_e$ and ${\cal Q}_e$, respectively, and both are coupled to
the scalar field $\Psi$, besides being coupled to each other.
Moreover, $A_i$ couples to $\Psi$ through kinetic terms, whereas
$\A_i$ couples through potential terms. Both gauge charges can be
obtained by the Gauss law, adapted to non-asymptotically
flat stationary spacetimes  and to the presence of the
dilaton \cite{by1,creighton-mann} (see also \cite{lemos1,lz1})
\begin{equation}
{Q_e} ={L_3\over 4\pi}\int_{{\cal S}_1}{\delta E_in^{i}}d{\cal S} \,,
\label{echarge1}
\end{equation}
\begin{equation}
{\cal Q}_e  = {L_3\over 4\pi} \int_{{\cal S}_1}{\delta \E_{i} n^{i}}d{\cal
S} \label{echarge1b} \, ,
\end{equation}
where $E_i\equiv e^{(\beta_1-\beta_0)\phi}F_{ij} t^{j}$, $\E_i\equiv
e^{(3\beta_1-\beta_0)\phi}\F_{ij} t^j/4$, $n^i$ is the unit normal to
the spacelike one-dimensional surface ${\cal S}_1$, a circle, and $t^{i}$ is
the timelike normal to the two-space ($D_2$), a disk.  Here also ${\cal S}_1$
is the infinite border of $D_2$.  As in the definition of mass and angular
momentum,  the symbol
$\delta$ indicates the difference between the quantity in question in the
considered spacetime and the same quantity in a reference spacetime. Namely,
$\delta E_i= E_i-(E_i)_o$ and $\delta\E_i = \E_i - (\E_i)_o$. Quantities
$E^i$ and $\E^i$ can be interpreted as electric fields in the two-space
orthogonal to $t^i$.  $(E_i)_o$ and $(\E_i)_o$ are the electric field
strengths for the background (or reference) spacetime, when no
localized objects are present. The integrals
in Eqs. (\ref{echarge1}) and (\ref{echarge1b}) are taken at spatial infinity.
Hence, in order to obtain a
well defined charge in spacetimes asymptotically
AdS, we must subtract the background value from the
corresponding global charge, and this procedure has to be
applied to every charge of the model.

It is worth mentioning that both of the electric charges are built
from conserved currents $J_e^i \equiv \nabla_j
\left(e^{(\beta_1-\beta_0)\phi}F^{ij}\right)$ and ${\cal J}_e^i
\equiv \nabla_j \left(e^{(3\beta_1-\beta_0)\phi}\F^{ij}/4\right)$,
respectively, for which follow immediately $\nabla_i J_e^i=0$ and
$\nabla_i{\cal J}_e^i =0$ (see also the item ({\it c}), in this
section,  below). These two identities guarantee the existence of
the two conserved electric charges as defined above.

\vskip 0.3cm

{\it (b) Magnetic charges of the gauge fields $A_i$  and
$\A_i$.}
The magnetic charges in 3D spacetimes have world histories with
dimension zero. They are events in the spacetime (instantons)
\cite{teitel,pisarski} since the
dual field strength in 3D is not a 2-form, but a 1-form,
i.e., $^{*}\!{\bf F}$ yields the 1-form $B_i
= \epsilon_{ijk}F^{jk}/2$. Thus, integration over the boundary of
$D_2$ at infinity cannot be performed in the same way as one does for
the electric charge. However, we can think of  the divergence $\nabla_i
B^i$ as the magnetic charge density and an integration over the
spacetime 3D world volume $\Sigma$ yields the magnetic charges (or magnetic
instantons). For stationary spacetimes, the divergence
$\nabla_i B^i$ defines an invariant charge density $\rho_m$ on the two-space
$D_2$,  and the volume integration over the whole $D_2$ yields the magnetic
charge. Using Gauss theorem, the volume integral over $D_2$ is changed into a
surface integral over the infinite boundary of $ D_2$, i.e., over the ${\cal
S}_1$ circle mentioned above.  We can then define
the magnetic charges in 3D by
\begin{equation}
{Q_m} ={L_3\over 4\pi}\int_{{\cal S}_1}{\delta B_i n^i}d{\cal S} \,,
\label{mcharge1}
\end{equation}
\begin{equation}
{\cal Q}_m={L_3\over 4\pi} \int_{{\cal S}_1}{\delta{\cal B}_{i}n^i}d{\cal S}
\label{mcharge1b} \, ,
\end{equation}
where $B_i= e^{(\beta_1-\beta_0)\phi}\, \epsilon_{ijk}F^{jk}/2$, ${\cal
B}_i=e^{(3\beta_1-\beta_0)\phi}\,\epsilon_{ijk}\F^{jk}/8$, and the
integration is taken over a ${\cal S}_1$ surface, at the spatial infinity of
$D_2$ (see above). These definitions apply at least for stationary
spacetimes and do not include the ``vortex magnetic charge" as
defined by some authors, where the static magnetic field can be
interpreted as being produced by a stationary electric current
(vortex) (see \cite{Hirschwelch,dias2}). They
 are certainly useful in the case of
instanton monopoles as defined in \cite{teitel,pisarski,helayel}.
Such definitions use the fact that, in 3D,  the monopole
generates a tangent electric field which can be used to determine the
magnitude of the charge \cite{helayel} (see Sect. \ref{secttoroidal}).
Let us emphasize that the surface integrals in Eqs. (\ref{mcharge1}) and 
(\ref{mcharge1b}) were obtained, using Gauss theorem, from a volume integration
over the whole space $D_2$. Therefore, the magnetic charges defined in such a
way are meaningful only if the surface integration is taken over the infinite
boundary of $D_2$. 
For another discussion on the difficulties to define  quasilocal gauge charges
associated with 4D dyonic black holes  see Ref. \cite{booth-mann}.

\vskip 0.3cm

\noindent {\it (c) Deformations of the electromagnetic harges.}
An investigation on the field equations shows that the electromagnetic
charges defined above may have additional contributions from the
interaction terms with other fields.
This is particularly true when $\Psi$ is nonzero, and the interaction
terms between the gauge field $\A_i$ and the scalar $\Psi$ give rise
to source terms in the field equations, in such a way that the
conserved currents acquire extra terms that depend on $\Psi$ and on ${\A}$.
To be more explicit, let us show what happens, for instance, regarding 
the electric charge $Q_e$. The full conserved current corresponding to
the ${\bf F}$ field is
\begin{equation}
J^{j}= \nabla_{i}\left[e^{(\beta_1-\beta_0)\phi}\left(F^{ij}
+F_{\rm extra}^{ij}\right)\right]\, ,
\end{equation}
where $F_{\rm extra}^{ij}\equiv \A^j\nabla^i\Psi -\A^i\,\nabla^j\Psi$.
Therefore, one  should add a
second term to Eq. (\ref{echarge1}) to give
$\displaystyle{\int_{{\cal S}_1}{\delta\left(E_{i}
+E^{\rm extra}_i\right)n^{i}}d{\cal S}}$, where $E^{\rm extra}_i =
\displaystyle{e^{(\beta_1-\beta_0)\phi}F^{\rm extra}_{ij}t^j}$.
 There are analogous correction terms
related to the other electric charge, ${\cal Q}_e$, and also to the magnetic
charges $Q_m$ and ${\cal Q}_m$.
However, we find that for the black hole solutions we are
going to analyze in this paper the above
mentioned corrections to  the conserved charges are zero.
The extra terms, in fact, contribute to the quasilocal charges, when the
boundary of integration ${\cal S}_1$ is not at infinity, but vanish at the
infinite boundary of $D_2$.

\subsubsection{Dilatonic charges}

In addition to their mass and electromagnetic gauge charges, stationary
asymptotically AdS 3D black holes are also
characterized by the dilaton charge. As a matter of fact, two dilaton
charges can be defined. The charge $Q_{\phi}$ \cite{ghs} and its dual
${\tilde Q}_{\phi}$ \cite{gibbonsperry} are given by
\begin{equation}
Q_{\phi}=
{L_3\over 4\pi}\int_{{\cal S}_1}{\delta\left(e^{(\beta_0+\beta_1)\phi}
\nabla_i\phi\right)n^i}d{\cal S} \, , \label{dilcharge}
\end{equation}
\begin{equation}
{\tilde Q}_{\phi} =
{L_3\over 4\pi}\int_{{\cal S}_1}{\epsilon_{ijk}\delta\left(
e^{(\beta_0+\beta_1)\phi}\nabla^k\phi\right) n^i t^j }d{\cal S} \, ,
\label{dilchargeb}
\end{equation}
where the integrations are defined in the same way as before.

The dilaton charge (\ref{dilcharge}) is defined in Ref. \cite{ghs}
in  the Einstein frame where $\beta_0+\beta_1=0$, and for $\Lambda \neq 0$
 it is not
related to a conserved current. It represents the total flux of
the vector field $V_i=\nabla_i\phi$ across the surface ${\cal S}_1$ at the 
boundary of the
space $D_2$. The result for $Q_\phi$ is the same to all stationary
observers at infinity  of $D_2$, and can then be identified with
the charge of the dilaton field. It is a conserved charge just in the 
case  $\Lambda=0$.
Although this surface integral 
does not come from a conserved current, and perhaps should not 
be called a charge, we maintain it here because sometimes it
has a nonzero value and is not totally useless 
(see \cite{ghs,kalloshetal}).

The dual dilaton charge (\ref{dilchargeb}), on the other hand,
is a conserved charge, since it is obtained from the conserved current
${\tilde J}_\phi^i= \nabla_j\left(\epsilon^{ijk}
e^{(\beta_0+\beta_1)\phi}\nabla_k\phi\right)$, which is divergence free
$\nabla_i {\tilde J}_\phi^i =0$.

\subsubsection{Charges of the scalar field $\Psi$}

Finally, due to the presence of the scalar field $\Psi$ in the
action (\ref{lagrN}) two other charges  can also be defined. {F}rom
Eq.$\,$(\ref{psieom}) it is possible to identify the quantity
$J_{\Psi}^i = e^{(\beta_0-\beta_1)\phi}\nabla^i\Psi + e^{(\beta_1-\beta_0)\phi}
\left(F^{ij}\A_j +\A^2\nabla^i\Psi -A^iA^j\nabla_j\Psi\right)$ 
as a conserved current, $\nabla_i J_\Psi^i =0$. 
The  corresponding
conserved  charge (analog to the electric charge)  is
\begin{equation}
Q_{\Psi}=
{L_3\over 4\pi}\int_{{\cal S}_1}
\delta\left(J_{\Psi}^{i}\right)n_{i}\,d{\cal S},\label{qpsi}
\end{equation}
where the integration and the symbol $\delta$ have the same meaning as 
above.  Let us mention that,  
for the solutions we are going to analyze here, the 
interaction terms
between $\Psi$ and the gauge fields ${\bf A}$  and $\cal A$ do not
contribute to the total charge for $\Psi$, and this charge is
identically zero.

It is also possible to define a second conserved charge of topological
character (which is the analog of the magnetic charge), also a source
of the scalar field $\Psi$. {From} the
vector quantity $\nabla^i\Psi$ we may construct an anti-symmetric dual
tensor as $H_{ij} = \epsilon_{ijk}\nabla^k\Psi$. Therefore, the vector
density $\nabla_iH^{ij}$ is divergence free and can be interpreted as
a conserved current. Thus there is an associated conserved charge 
defined by
\begin{equation}{\tilde Q}_{\Psi} = {L_3\over
4\pi}\int_{{\cal S}_1}{\epsilon_{ijk}\delta\left(\nabla^k\Psi \right)\, t^i
n^j}d{\cal S} \, ,
\label{qpsidual}
\end{equation} where the integration is
the same as defined above.  When ${\cal A}\neq 0$, there are
additional terms in the equation of motion for $\Psi$ not considered
to arrive at Eq. (\ref{qpsidual}), but they do not contribute to the conserved
charge [see the comments just after Eq. (\ref{mcharge1b})].

\subsubsection{General comments }

We  now study explicitly the connection between the general 4D
stationary asymptotically AdS spacetimes and the corresponding 3D
metrics obtained through the dimensional reduction technique
discussed in the present section. We will reduce through the
angular coordinate $\varphi$. We study 4D toroidal black
holes in AdS spacetime, charged and rotating. These have a
straightforward dimensional reducing procedure; the theory
obtained is a 3D Brans-Dicke theory, and the 3D results are
clear cut and simple; most frames (including the Einstein frame)
are good frames.

\section{3D Charged Rotating Toroidal Black Holes}\label{secttoroidal}

\subsection{The 4D metric and parameters}

Hereafter we consider a class of rotating black holes with
toroidal topology.
In a previous paper \cite{lz1} we reported a rotating electrically
charged black hole with a toroidal horizon.  Before going on to the
dimensional reduction of such a black hole, it is worth noting that
following the same procedure as in \cite{lz1} a dyonic version of the
black hole can be found.

We start constructing the static dyonic toroidal black hole by choosing
the coordinate system $({t}, r,\theta, {\varphi})$ with
$-\infty<{t}<+\infty$, $0\leq r< +\infty$, $0\leq\theta<2\pi$,
$0\leq{\varphi}< 1$ (the ranges of the angular coordinates are
arbitrary, we have chosen these particular ones to yield convenient
values for the mass and charges).

The solution is found by solving Einstein-Maxwell equations for such a
static   spacetime. We find 
\begin{eqnarray}
d\hat s^{2}&=&-\left(\asr-\frac{4m}{r} +\frac{4(q^{2}+g^2)}{r^2}
\right)d{{t}}^{2} +\frac{dr^{2}}{\asr-\frac{4m}{r}
+\frac{4(q^{2}+g^2)}{r^2}}\nonumber \\ & &
+r^{2}\left(d{{\varphi}}^{2}+d\theta^{2}\right)\, ,
                          \label{cylstat} \\
{\bf \hat A }&=&-{2q\over  r}\,{\rm\bf d}t
 -  {2g}\,\theta {\rm\bf d}\varphi\, ,
                          \label{gaugest}
\end{eqnarray}
where $\alpha^2 \equiv -\frac{1}{3}\Lambda$, and $m$, $q$ and $g$ are
integration constants.  It is easy to show, for instance using Gauss
law, that $q$ and $g$ are respectively the electric and magnetic
charges of the black hole, and $m$ is its mass.  Depending on
the relative values of $m$, $q$ and $g$, metric (\ref{cylstat}) can
represent a static toroidal black hole.  

The rotating metric is then
obtained by performing a local coordinate transformation which mixes
time and angular coordinates. The result can be written in the form
\begin{eqnarray}
d\hat s^{2}& =& -\left(1-\frac12 a^2\alpha^2\over 1-\frac32 a^2\alpha^2\right)
\left[{\Delta\over r^2}\left(dt -
{a\over\sqrt{1-\frac12 a^2\alpha^2}}d\varphi\right)^2 +r^2\left(\,d\varphi
-{a\alpha^2\over\sqrt{1-\frac12 a^2\alpha^2}}\,dt\right)^2\right] 
\nonumber \\ & &
+ r^2\left(\frac{dr^2}{\Delta} +d\theta^2\right)\, ,
\label{cylmetr}\\
{\bf \hat A} & =& -2 {q\over r}\left({\rm\bf d}t -
{a\over\sqrt{1-\frac12 a^2\alpha^2}}\,{\rm\bf d}\varphi\right) -
2{g}\theta\left({\rm\bf d}\varphi -
{a\alpha^2\over\sqrt{1-\frac12 a^2\alpha^2}}\, {\rm\bf d}t\right)\, ,
\label{cylgauge}
\end{eqnarray}
where
\begin{equation}
\Delta =\alpha^2r^4-4m\,\left(1-\frac32 a^2\alpha^2\right)\,{r}
 + 4\left(q^2+g^2\right)\left(\frac{1-\frac32 a^2\alpha^2}{1-\frac12
a^2\alpha^2}\right)\, . \\ \label{cyldelta}
\end{equation}
Parameters $m$, $q$ and $g$ have the same interpretation as in the static
black hole (\ref{cylstat}).
The rotation parameter $a$ is
defined through
$\displaystyle{J=\frac{3}{2}aM \sqrt{1-{a^2\alpha^2}/{2}}\,}$,
where $J$ is the angular momentum of the black hole. Let us also mention that
the above choice of parameters, with the constraint $0\leq a^2\alpha^2\leq
1$,  ensures that the asymptotic form of the metric  for large $r$  is
exactly the static AdS metric (see Ref. \cite{lz1}). In the following,
however, we restrict the analysis to the case $0\leq a^2\alpha^2 <{2\over 3}$.

The metric (\ref{cylmetr}) admits two spacelike Killing vectors, so
that two independent dimensional reductions are allowed in this case.
Such a metric can then be reduced from 4D to two different 3D black
hole solutions, or from 4D to one 2D nontrivial black hole. We are
going to consider the reduction along $\partial_\varphi$. For the
reduction along the other Killing vector ${\partial_\theta}$ the
result is the dyonic analog of the 3D rotating charged black hole
obtained in \cite{lz1} (see also \cite{lemos1}), in which case one has
to consider a different gauge for ${\bf \hat A}$, namely,
${\bf \hat A}= -2 {q\over r}\left({\rm\bf d}t -
{\left(a/\sqrt{1-{a^2\alpha^2}/2\,}\right)}\,{\rm\bf d}\varphi\right) -
2{g}\left(\varphi
-{\left(a\alpha^2/\sqrt{1-{a^2\alpha^2}/2}\,\right)}\,t\right) {\rm\bf
d}\theta$,  without changing the electromagnetic Maxwell field.

\subsection{The 3D black hole spacetime }

\subsubsection{ The metric and charges}\label{sect3Dtor}

Using the prescriptions developed in Sect. II,
the dimensional reduction along the Killing direction $\varphi$ can now be
performed, yielding the following 3D static black hole
\begin{eqnarray}
e^{2\beta_0\phi}ds^{2}& =&
-{\left(1-\frac32{a^2\alpha^2}\right)\,\Delta\over
r^2(1-\frac12{a^2\alpha^2})-{a^2\Delta\over r^2}}\, dt^2
+ r^2\left(\frac{dr^2}{\Delta} +d\theta^2\right)\, , \label{cylmetrred1}\\
{\bf A} & =& -2 \left({q\over r}\,
-g\,{a\alpha^2\over\sqrt{1-\frac12{a^2\alpha^2}}\,}\,
\theta\,\right) {\rm{\bf d}}t\, ,\label{cyl3Dgauge}\\
{\bf {\cal A}} &=& {a\left(\Delta - \alpha^2r^4\right)
\over  r^4(1-\frac12{a^2\alpha^2}) - a^2 \Delta}\,L\,
{\rm \bf d}t\, , \label{cyl3Dkkgauge} \\
L^2e^{2\beta_1\phi} &=&{1\over 1-\frac32{a^2\alpha^2}}\left[r^2
\left(1-\frac12{a^2\alpha^2}\right)-{a^2\Delta\over r^2} \right]\,,
\label{cyldilaton}\\
\Psi&=& {2 q \over r}{a\over\sqrt{1-\frac12{a^2\alpha^2}\,}}
-2\,g\,\theta\, ,
\label{cyl3Dpsi} \end{eqnarray}
where  the new  arbitrary constant $L$ introduced in the
definition of the dilaton (\ref{cyldilaton}) carries physical dimensions of
length.

For $q^2+g^2 \leq {3\over 4}(1-\frac12{a^2\alpha^2})\sqrt[3]
{(1-\frac32{a^2\alpha^2})\,m^4/\alpha^2\,}\,$, the above solution
represents a static spherically symmetric three-dimensional black
hole with electric and magnetic gauge charges proportional to $q$
and $g$, respectively, and with an extra gauge charge proportional
to $a\sqrt{1-\frac12{a^2\alpha^2}}$.  For $q^2+g^2 > {3\over
4}(1-\frac12{a^2\alpha^2})\sqrt[3]
{(1-\frac32{a^2\alpha^2})\,m^4/\alpha^2\,}$ the above solution
represents a naked singularity, with singularities at points where
$ r^4(1-\frac12{a^2\alpha^2}) - a^2\Delta=0$. For future
reference, and to remind of the toroidal topology of the original
4D black hole,  we call the above solution as the 3D toroidal
black hole. One should keep in mind, however, that the topology of
the 3D solution (\ref{cylmetrred1}) is in fact spherical, or, more
precisely, circular, because the slices $t=$constant are
two-dimensional spacelike surfaces.

For the sake of definiteness we choose initially the good frame
$\beta_0=0$.
Using the definition of Sect. \ref{sectdefs} we can now determine the
mass and charges of the present solution.  In order to apply
Eq. (\ref{massdef}) to calculate the mass of the toroidal 3D black
hole, let us firstly define explicitly the quantities appearing in
that equation.  In metric (\ref{cylmetrred1}) we then choose a region
$\cal{M}$ of spacetime bounded by $r={\rm {\rm constant}}$, and two
space-like hypersurfaces $t=t_{1}$ and $t=t_{2}$. The hypersurface
$t={\rm constant}$, $r={\rm constant}$, is the one-dimensional
boundary ${\cal S}_1$ of the two-space $D_2$. The boundary of
$\cal{M}$, $\partial \cal M$, in the present case consists of the
product of ${\cal S}_1$ with timelike lines ($r={\rm constant }\,
,\theta={\rm constant}$) joining the surfaces $t=t_1$ and $t=t_2$, and
these two surfaces themselves.  ${\cal S}_1$ can also be thought as
the intersection of $D_2$ with $\partial \cal M$ (${\cal S}_1$ is a
circle with radius $r$).
 The induced metric $\sigma_{ab}$ is obtained from (\ref{cylmetrred1})
by putting $dt=0$ and $dr=0$. Thus, $a,b=2$ and
$\sigma_{ab}=\sigma_{22}\equiv\sigma= r^2$, while the
two-space metric $h_{ij}$ is obtained by putting $dt=0$.

Using Eqs. (\ref{kappadef}) and (\ref{cylmetrred1}), we get the following
expression for the extrinsic curvature of ${\cal S}_1$,
\begin{equation}
 {k^\phi}= -{e^{-\beta_1\phi}\over 2}{\sqrt{\Delta} \over r}
\,\,\left({2\over r} +
2 {\partial \left(\beta_1\phi\right)
\over \partial r}\right)\, , \label{cylkappa}
\end{equation}
where $\Delta$ and $e^{\beta_1\phi}$ are given respectively
by (\ref{cyldelta}) and (\ref{cyldilaton}).
To build $\delta\left(k^\phi\right)$ we compute $k^\phi$ from the
full solution given in Eqs. (\ref{cylmetrred1})--(\ref{cyl3Dpsi}),
that describe the 3D toroidal black hole in an asymptotically
AdS spacetime. The background
spacetime is the 3D spherical AdS spacetime with no black hole
present, whose  extrinsic curvature $\left(k^\phi\right)_{\!o}$ follows
from the same metric (\ref{cylmetrred1}) by choosing $m=0$, $q=0$ and $g=0$.
Then, using (\ref{cylkappa}) to calculate  $\delta\left(k^\phi\right)=
k^\phi- \left(k^\phi\right)_{\!o}$ and substituting
into (\ref{massdef}), and taking the limit $r\longrightarrow\infty$,
the mass of the toroidal 3D black hole is finally obtained,
\begin{equation}
M_{{\rm 3D}} = m\left(1 +a^2 \alpha^2\right) \,
,\label{masstor} \end{equation}
where $m$ is the mass of the 4D black hole, and to simplify we have put
$L=L_3$. We see that the reduced 3D
black hole acquires mass from the original toroidal black hole in 4D
spacetime.  The additional mass, $\delta M=m\,a^2\alpha^2$, depends
explicitly on the 4D rotation parameter $\omega=a\alpha^2/\sqrt{1-\frac32
a^2\alpha^2}$, and can be viewed as being generated by the motion of the 3D
system along the extra dimension. That is to say, the same well known
mechanism that gives rise to the electromagnetic field and charges in
Kaluza-Klein theories, also gives rise to part of the mass of the system in the
compactified spacetime.

As  mentioned before, metric (\ref{cylmetrred1}) is static and
the angular momentum is zero. This  can be seen using Eq.
(\ref{momentumdef}) which gives,
\begin{equation}
J_{3D} =0 \, .
\end{equation}

According to Eq. (\ref{cyl3Dgauge}), both the 4D electric and magnetic
charges ($q$, $g$) are sources to the 3D electromagnetic field. When
considered as independent sources, $q$ and $g$ generate electric
fields with distinct geometric properties. $q$ is the source of a
radial field $E_r \sim {q\over r^2}$, while $g$ gives rise to a
tangent (uniform) electric field $E_\theta\sim
ga\alpha^2/\sqrt{1-\frac12{a^2\alpha^2}}$.  On the other hand, the
source for the Kaluza-Klein gauge field  ${\cal A}$ is
proportional to the 4D rotation parameter $\omega=
a\alpha^2/\sqrt{1-\frac32{a^2\alpha^2}}$.
 Such a charge generates a radial electric field.
Electric gauge charges for $\bf A$ and $\cal A$ for the 3D
toroidal black hole are obtained
from Eqs. (\ref{echarge1}) and (\ref{echarge1b}),  and are
given respectively by
\begin{equation}
{ Q}_e =  q \, ,
\label{toroielectr1}
\end{equation}
and,
\begin{equation}
{\cal Q}_e =\frac32 ma\sqrt{1-{1\over2} a^2\alpha^2} =  J \, .
\label{toroielectr2}
\end{equation}
The electric  charge ${\cal Q}_e$, source to the
Kaluza-Klein gauge field,
is proportional to the 4D angular momentum $J$, as expected.

 Magnetic gauge charges are calculated from Eqs.
(\ref{mcharge1}) and (\ref{mcharge1b}). Following the same prescription
as for calculating the other gauge charges we find
\begin{equation}
{ Q}_m ={ga\alpha\over \sqrt{1-{a^2\alpha^2\over2}}}\, ,
\label{toroimagnetic1}
\end{equation}
and,
\begin{equation}
{ \cal Q}_m = 0\, .
\label{toroimagnetic2}
\end{equation}

Thus,  we see that the rotation of the 4D
charged black hole generates a stationary magnetic current $i_g=
ga\alpha^2/\sqrt{1-\frac12{a^2\alpha^2}}$ which is the source of a
tangent electric field. In the dimensionally reduced static 3D
spacetime, there is no frame dragging and the tangent electric field
must be generated by a Dirac monopole, whose magnitude of the charge
is ${Q}_m = i_g/\alpha$. The existence of a uniform tangent electric
field is a special feature of associated to the presence of a Dirac
monopole in 2+1--dimensional spacetimes \cite{helayel}.

The dilaton charges ${ Q}_\phi$ and $\tilde
Q_\phi$ for the toroidal black hole are both zero,
\begin{equation}
Q_\phi = 0\, , \hskip 1.5cm  \tilde Q_\phi = 0\, .
\end{equation}
It is worth noticing, however, that
$\delta\left(e^{(\beta_0+\beta_1)\phi}\right)$ in Eqs. (\ref{dilcharge}) and
(\ref{dilchargeb}) is nonzero and the quasilocal dilaton charges are both
nonzero and depend on  the surface of integration ${\cal S}_1$.  Consider,
for instance, the case of Eq. (\ref{dilcharge}) and let $Q_\phi(r)$ be the
quasilocal charge obtained for  $\phi$  when the integration  boundary
${\cal S}_1$ is at $r=$constant (not at infinity). $\delta(e^{\beta_1\phi})$
is  the difference between the full dilaton field
 $e^{\beta_1\phi}$ ($\beta_0=0$),  given by  (\ref{cyldilaton}), and  the
background  dilaton field  (from the  background spacetime)
$\left(e^{\beta_1\phi}\right)_{\!o}$,
which follows from (\ref{cyldilaton}) by putting $m=0$, $q=0$,
$g=0$.
The resulting conserved dilaton charge can then be thought as
the asymptotic limit ($r\longrightarrow \infty$)  of
the quasilocal charge
$$
Q_\phi(r) = { 3\over 2}{m a^2\alpha\over r}\, ,
$$
where as mentioned before we choose $\beta_1 = -2$.
Hence, when the  integration is taken over the infinite boundary,
the total charge $Q_\phi= \lim_{r\longrightarrow\infty}
Q_\phi(r)$ vanishes.  Similar
arguments hold for the  dual charge ${\tilde Q}_\phi$.
This result can be interpreted as the
dilaton being a short range field.

We now investigate the physical meaning of the scalar field $\Psi$ by
firstly calculating its charges. In order to do that we substitute  Eq.
(\ref{cyl3Dpsi}) into   Eqs. (\ref{qpsi}) and (\ref{qpsidual}) and
take the appropriate limit to get,
\begin{equation}
Q_\Psi = 0\, , \hskip 1.5cm  \tilde Q_\Psi = g\, .
\end{equation}
Eq. (\ref{reducedA}) tells us that $\Psi$ is a gauge field, which
appears in 3D as a scalar field.  The nonzero charge
$\tilde Q_\Psi$, the source of a scalar field, is the analog of the magnetic
charge and can be viewed as a topological charge.

The solution given by (\ref{cylmetrred1}) then represents a localized
nonrotating massive object in an asymptotically AdS spacetime, and has
three gauge charges ($Q_e$, ${\cal Q}_e$ and $Q_m$) and one
scalar charge $\tilde Q_\Psi$.

\subsubsection{Singularities, horizons, and causal structure}

The 3D spacetimes here derived are circular and
static. Their causal structure  is in several aspects similar
to the causal structure of the corresponding 4D spacetime.
The main important difference between 4D and 3D
solutions is related to the singularities, as it can be seen by
comparing the respective curvature invariants. In the 4D spacetime
(Eq. (\ref{cylmetr})) there is a singularity at $r=0$, whilst in 3D
there are other points where the curvature is singular.  This is
verified by studying the 3D Ricci and Kretschmann scalars of metric
(\ref{cylmetrred1}), which are given respectively by
\begin{eqnarray}
R& =& -26\alpha^2 +{8m\over r^3}\left(1-\frac32{a^2\alpha^2}\right) +
{8(q^2+g^2)\over  r^4}\left(\frac{1-\frac32{a^2\alpha^2}}
{1-\frac12{a^2\alpha^2}}\right) +12\alpha^2 {\Delta \over \Gamma}+
{3\over 2}{\Gamma'\Delta' \over r^2 \Gamma}+\nonumber\\
 & & -{3\over2}{\Delta\Gamma'^2\over r^2 \Gamma^2}+
2{\Delta\Gamma'\over r^3\Gamma}\, , \\
 K  &= &364\alpha^4-128{\alpha^2\,m\over r^3}
\left(1-\frac32{a^2\alpha^2}\right) -224{(q^2+g^2)\alpha^2\over
r^4}\left(\frac{1-\frac32{a^2\alpha^2}} {1-\frac12{a^2\alpha^2}}\right)
+\nonumber \\
 & & +160{m^2\over r^6}\left(1-\frac32{a^2\alpha^2}\right)^2
-256{m\over r^7}{(q^2+g^2) \over \left(1-\frac12{a^2\alpha^2}\right)}
\left(1-\frac32{a^2\alpha^2}\right)^2 +\nonumber\\
& & + 192{(q^2+g^2)^2\over r^8}\left(\frac{1-\frac32{a^2\alpha^2}}
{1-\frac12{a^2\alpha^2}}\right)^2 -288\alpha^4 {\Delta\over \Gamma}
-36\alpha^2 {\Delta'\Gamma'\over r^2\Gamma} -48\alpha^2 {\Delta\Delta'\over
r^3 \Gamma} - 24\alpha^2 {\Delta\Gamma'\over r^3\Gamma}  +\nonumber\\
 & & +48\alpha^2{\Delta^2 \over r^4 \Gamma} -6{\Delta'^2\Gamma'^2\over r^5
\Gamma}+ 144\alpha^4{\Delta^2\over \Gamma^2}
+36\alpha^2{\Delta\over r^2} \left({\Gamma'^2\over\Gamma^2}+
{\Delta'\Gamma'\over \Gamma^2}\right) +24\alpha^2{\Delta^2\Gamma'\over
r^3\Gamma^2} +{9\over4} {\Delta'^2\Gamma'^2\over r^4\Gamma^2} + \nonumber\\
& & +
9{\Delta\Delta' \Gamma'^2\over r^5\Gamma^2}  -4 {\Delta^2\Gamma'^2\over r^6
\Gamma^2}+ -36\alpha^2{\Delta^2\Gamma'^2\over r^2\Gamma^3}
-{9\over2}{\Delta\Delta'\Gamma'^3\over r^4\Gamma^3} -3{\Delta^2\Gamma'^3\over
r^5\Gamma^3} +{9\over4}{\Delta^2\Gamma'^2\over r^4\Gamma^4} \, ,
\end{eqnarray}
where we have defined $\Gamma = \left(r^4(1-\frac12{a^2\alpha^2}) -
a^2\Delta\right)/(1-\frac32{a^2\alpha^2})$ and
$' \equiv {\partial\over\partial r}$.  The 3D spacetime
(\ref{cylmetrred1}) then shows singularities  when the  following condition
is  fulfilled
\begin{eqnarray}
& &r^2\left[r^4 + {4m \,a^2 } r
-4(q^2+g^2)\,{a^2\over 1-\frac12{a^2\alpha^2}}\right]=0\, .
\label{cylsing}  \end{eqnarray}
The expression among brackets in Eq. (\ref{cylsing}) has, for
$q^2+g^2\neq 0$, one positive (real) root $r_s$
signaling the presence of a singularity. It has also a negative
root which we do not consider. Thus, infalling geodesic particles
coming from large $r$ hit a singularity at $r=r_s$ where the spacetime
ends (see Fig. 1).
\begin{figure}
\begin{center}
\mbox{\epsfig{figure=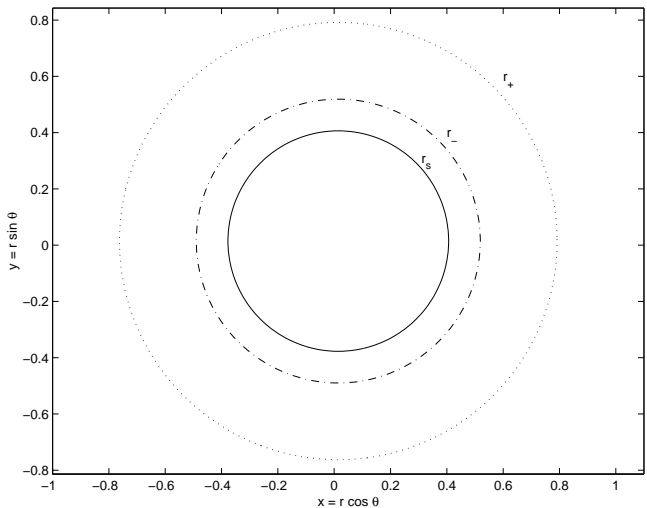,height=3.01in}}  \label{torsingfig}
\end{center}
\caption{The singularity  and horizons of the toroidal 3D space-time in the
coordinates of metric (\ref{cylmetrred1}).  The parameters are such that the
event horizon at $r_+$ (dotted line) and Cauchy horizon at $r_-$
(slash-dotted line) are present. Spacetime ends at
the singularity $r_s$ (full line).}
\end{figure}

The horizons of (\ref{cylmetrred1}) are given by the real
roots of the equation
\begin{eqnarray}
& &\alpha^2 r^4 - 4m\left(1-\frac32{a^2\alpha^2}\right) r +
4(q^2+g^2)\left(\frac{1-\frac32 a^2\alpha^2}{1-\frac12 a^2\alpha^2}
\right) =0\, .
\label{cylhor}
\end{eqnarray}
In the analysis of horizons, the relevant function is
\begin{equation}
 Descr=   {3\over
4} (1-\frac12{a^2\alpha^2})\sqrt[3]
{(1-\frac32{a^2\alpha^2})\,\frac{m^4}{\alpha^2}\,} -q^2 -g^2
\end{equation}
Depending on the relative values of mass $m$, charges $q$, $a$ and
$g$, we have three distinct cases to analyze
(the case $q^2+g^2=0$ is considered in next subsection):
(i)  $Descr>0$ -- in such a case the
metric (\ref{cylmetrred1}) has two horizons, the event horizon at
$r_+$ and the Cauchy or inner horizon at $r_-$. This is shown in
Fig. 1.  The singularity at $r=r_s$ is enclosed by both horizons.
The spacetime can then be extended through the horizons till $r_s$.
It represents a static black hole. Note that the other region between
$r_s$ and $r=0$, belongs to a disconnected spacetime and we do not
analyze it further;
(ii) $Descr=0$ --
the solution is the extreme black hole spacetime. There is only one
horizon at $r =r_+=r_- = \sqrt[3]{m\left(1-\frac32
a^2\alpha^2\right)/\alpha^2}\,=\sqrt[4]{\frac43
\frac{(q^2+g^2)}{\alpha^2}\left(\frac{1-\frac32 a^2\alpha^2}{1-\frac12
a^2\alpha^2}\right)\,}$.  In such a case, when drawing an
analog figure to Fig. 1, the dotted and the dashed
(external) lines would coincide with each other, and the singularity at
$r=r_s$ (solid internal line) would be still hidden (to
external observers) by the horizon. Geodesic inward lines end at the
singularity $r_s$;  (iii) $Descr<0$ --  this solution has no horizons
and represents a naked singularity. From the above description,
the Penrose diagrams, with the inherent topology and causal structure 
of spacetime,  can easily be drawn.

\subsubsection{Special cases}

\noindent{\it (a) The $J=0$ case}

This case corresponds to an uncharged version of the solution studied
in Sect. \ref{sect3Dtor}, since $J=0$ implies $a=0$, what makes the
Kaluza-Klein electric charge ${\cal Q}_e$ equal to zero. The magnetic charge
$Q_m$ also vanish (see Eqs. (\ref{toroielectr2}) and (\ref{toroimagnetic1})).

The 3D metric and other fields are obtained also by the dimensional reduction
of the static charged 4D spacetime given by Eqs. (\ref{cylstat}) and
(\ref{gaugest}), which yield,
 \begin{eqnarray}
d s^{2}&=&-\left(\asr-\frac{4m}{r} +\frac{4(q^{2}+g^2)}{r^2}
\right)d{{t}}^{2} +\frac{dr^{2}}{\asr-\frac{4m}{r}
+\frac{4(q^{2}+g^2)}{r^2}} +r^{2}d\theta^{2}\, ,      \label{cylstat3D} \\
{\bf  A }&=&-{2q\over  r}\,{\rm\bf d}t \, ,\label{gaugest3D}\\
L^2e^{2\beta_1\phi} &=& r^2\, ,\label{dilatonest}\\
\Psi & =& -  {2g}\,\theta \, , \label{Psistat}
\end{eqnarray}
and the other fields vanish.
This solution represents a 3D  static  spherically symmetric charged 
black hole whose geodesic and causal structures are the same as the
$\varphi=$ constant plane of the 4D static toroidal black hole
(\ref{cylstat}). Such a black hole has mass $m$, electric charge $q$
and an additional charge $g$, the source to the $\Psi$ scalar field.
The dilaton charge is zero. The Ricci and Kretschmann curvature
scalars are, respectively, $R= -6\alpha^2 -{12(q^2+g^2)\over r^4}$ and
$K= 12\alpha^4 + {96m^2\over r^6}+ {32\alpha^2 (q^2+g^2)\over r^4}
-{512m(q^2+g^2)\over r^7} + {704\left(q^2+g^2\right)^2\over r^8}$, showing
that  there is a singularity at $r=0$. For
$q^2+g^2 \leq {3\over4} \sqrt[3]{ \frac{m^4}{\alpha^2}\,  }$ 
there are two horizons and
the singularity is hidden to asymptotic  external observers. 
On the other hand, if
$q^2+g^2 > {3\over4}\sqrt[3]{ \frac{m^4}{\alpha^2} \,}$ 
the singularity is naked.

\medskip

\noindent{\it (b) The uncharged $q^2 +g^2 =0$ case}

An interesting 3D spherical black hole is obtained by the
dimensional reduction of the 4D toroidal rotating uncharged
black hole.  Such a solution can also be obtained directly from
(\ref{cylmetrred1})--(\ref{cyl3Dpsi}) by putting $q=0$ and $g=0$.
Namely,
\begin{eqnarray}
ds^{2}& =& -{\alpha^2r^2-{4m\over r}\left(1-\frac32
a^2\alpha^2\right) \over 1+{4ma^2\over r^3}}\, dt^2 +\frac{dr^2}{\asr
-{4m\over r}\left(1-\frac32 a^2\alpha^2\right)}  +r^2\,d\theta^2\, ,
\label{cylmetrred2}\\
{\bf {\cal A}} &=& -{4m\,a\sqrt{1-\frac12 a^2\alpha^2}\over
r^3 + 4m\,a^2}\,L\, {\rm \bf d}t\, , \label{cylkkgauge2}\\
L^2e^{2\beta_1\phi} &=& r^2 +{4m\,a^2 \over r}
 \, , \label{cyldilaton2} \end{eqnarray}
with the other fields being identically zero.
This solution corresponds to a 3D charged static black hole with just
one gauge field whose electric charge is ${\cal Q}_e =\frac32 m
a\sqrt{1-{1\over2} a^2\alpha^2}$.  The mass and angular momentum are
the same as for the case $q^2+g^2\neq 0$.  Singularities and horizons
of this spacetime are also easily obtained from the solution studied
in Sect. \ref{sect3Dtor} with $q^2+g^2=0$.  For all possible values
of parameters $\alpha^2>0$, $m(1-\frac32 a^2\alpha^2)>0$ and
$a^2(1-\frac32 a^2\alpha^2)>0$, there is always just one horizon at
$r= \sqrt[3]{4(1-\frac32 a^2\alpha^2)\frac{m}{\alpha^2}\,}$, 
and a singularity
at $r=0$.
\medskip

\noindent{\it (c) The $\alpha=0$ case}

After dimensional reduction along the
$\partial_\varphi$ direction the metric and the other potentials
can be obtained directly by making $\alpha^2=0$ in Eqs.
(\ref{cylmetrred1})--(\ref{cyl3Dpsi}), which give
\begin{eqnarray}
ds^{2}& =&-{\left[{-4m\over r} +{4(q^2+g^2)\over
r^2}\right] \over 1 +{4ma^2\over r^3}-{4(q^2+g^2)\,a^2\over r^4} }\, dt^2
+ \frac{dr^2}{{-4m\over r}+{4(q^2+g^2)\over r^2}} + r^2 d\theta^2\, ,
\label{cyl3d0}\\
{\bf A} & =& -2 {q\over r}\,{\rm{\bf d}}t\, ,\label{cyl3Dgauge0}\\
{\bf {\cal A}} &=& {a\left[-4mr -4(q^2+g^2)\right]
\over r^4 +4m a^2r -4(q^2+g^2)\,a^2}\,L\,
{\rm \bf d}t\, , \label{cyl3Dkkgauge0} \\
L^2e^{2\beta_1\phi} &=&r^2+ {4m a^2\over r}-{4(q^2+g^2)a^2\over r^2}\, ,
\label{cyldilaton0}\\
\Psi&=& {2 q \over r}{a}-2\,g\,\theta\, . \label{cyl3Dpsi0}
\end{eqnarray}
We have two distinct cases.
 Indeed, if $q^2+g^2\neq0$ there is one
horizon, $r_h= (q^2+g^2)/m$. On the other hand,  if $q^2+g^2=0$ there
are no horizons and the singularity is naked: in contrast to the 4D
black holes with spherical horizons the 4D uncharged toroidal black
holes vanish when the cosmological constant is set to zero, leaving a
naked singularity.  In both cases, the asymptotic region
$r\longrightarrow \pm\infty$ is not well defined.  We do not comment
further on this case.

\medskip

\noindent{\it (d) The rotating black hole}

One may, if one wishes, put this black hole to rotate by
performing a forbidden coordinate transformation which
mixes time and angles. This yields a new rotating solution.

\subsection{The 3D black hole spacetime in other frames}

Up to now we have analyzed the 3D black hole in the good frame.
Once we have the metric in the
good frame, the metric in any other frame can be obtained by
the conformal transformation given in Eq. (\ref{conftransf}).
We consider first the Einstein frame which follows by putting
$(\beta_0)_{\rm o}= -(\beta_1)_{\rm o}$, or by using
Eqs. (\ref{cylmetrred1}) and  (\ref{cyldilaton}) and choosing
$\beta_0 =-\beta_1$. This gives,
\begin{equation}
ds^{2}_{\rm E}
= -\Delta\,d\bar t^2 +\left(1-\frac12 a^2\alpha^2\over 1-\frac32
a^2\alpha^2\right) \left(\alpha^2\bar r^4-
{a^2 \alpha^2\over 1-\frac12 a^2\alpha^2}\,\Delta\right)\left(\frac{d\bar
r^2}{\Delta} + d\bar \theta^2\right)\, ,\label{cyl3Deinst}
\end{equation}
where $\Delta$ is defined as before,
and all the other fields keep the same form of Eqs.
(\ref{cyldilaton})--(\ref{cyl3Dpsi}).
We have dressed the coordinates with bars to
make clear they are different from the  metric in the good
frame (\ref{cylmetrred1}). This solution and (\ref{cylmetrred1}) are
conformally equivalent, except in the loci  $r=0$ and $\Gamma(r)=
r^4(1-\frac12 a^2\alpha^2)-a^2 \Delta(r)= 0$, where the
 metric (\ref{cylmetrred1}) is singular.
Metric (\ref{cyl3Deinst}) presents horizons at points where
$\Delta(\bar r)=0$, and singularities when $\Gamma(\bar r) = \bar
r^4(1-\frac12 a^2\alpha^2)-a^2 \Delta(\bar r)=0$.
The charges for both of the metrics are also the same.
Good and Einstein frames in this case both  yield black holes.

Other frames  can also be considered. For comparison, we show also the metric
of the toroidal 3D black hole in the string frame. Once again, we start
with the metric in the good frame and use Eq. (\ref{conftransf}), where now
$\left({\beta_0\over\beta_1}\right)_{\!\rm o} = -(1 \mp\sqrt{2})$, yielding
\begin{eqnarray}
ds^2_{\rm s}&=& -\left[{ r^4\left(1-\frac12 a^2\alpha^2\right)
  - a^2 \Delta \over 1-\frac32 a^2\alpha^2}\, \right]^{\mp\sqrt{2}}
\Delta\,d t^2 +\nonumber\\
& & \left[{ r^4\left(1-\frac12 a^2\alpha^2\right)
- a^2\Delta  \over 1-\frac32 a^2\alpha^2}\,
\right]^{1\mp\sqrt{2}}\left(\frac{d  r^2}{\Delta} + d\theta^2\right)\, ,
\label{cyl3Dstring} \end{eqnarray}
 and all the other fields keep the same form of
Eqs. (\ref{cyldilaton})--(\ref{cyl3Dpsi}).  The string and Einstein frames
are related by $
ds^2_{\rm s}= 
\left(e^{-\phi}\right)_{\!{\rm E}}^{\mp\sqrt{2}}ds^{2}_{_{\rm E}}$,
where $\left(e^{2\beta_1\phi}\right)_{\!{\rm E}}$ is the dilaton field
in the Einstein frame where $\beta_1=-1/2$.  Properties of the metric
in the string and Einstein frames are very similar, with the same
singularities and horizons. The 
conformal transformation relating the two frames is well defined everywhere
except at points where $r^4(1-\frac12 a^2\alpha^2)-a^2 \Delta(r)= 0$, which
correspond to singularities of the spacetime.

\subsection{The 2D black hole spacetime}

One can reduce one more dimension. Our
aim is now to go from the 3D  black hole presented in Eqs.
(\ref{cylmetrred1})--(\ref{cyl3Dpsi}) to the corresponding 2D reduced
black hole by performing a consistent truncation along the
$\theta$ direction.
In order to perform a consistent dimensional reduction
we have to choose $g=0$ in Eqs. (\ref{cyl3Dgauge}) and (\ref{cyl3Dpsi}).
The result is a
(1+1)--dimensional black hole solution of gravity theory with two gauge
fields $\bf A$ and $\cal A$,
two dilaton $\phi_1$ and $\phi_2$, and with one scalar
field $\Psi$:
  \begin{eqnarray}
 e^{2(\beta_0\phi_1+\bar\beta_0\phi_2)}ds^{2}& =&
-{\left(1-\frac32{a^2\alpha^2}\right)\,\Delta\over
r^2(1-\frac12{a^2\alpha^2})-{a^2\Delta\over r^2}}\, dt^2
+ r^2\frac{dr^2}{\Delta}\, ,\label{twometr}\\
 {\bf A} & =& -2 {q\over r}\, {\rm\bf d}t \, , \\
{\bf {\cal A}} &=& {a\left(\Delta - \alpha^2r^4\right)
\over r^4(1-\frac12{a^2\alpha^2}) - a^2\Delta}\,L\,
{\rm \bf d}t\, , \label{twokkgauge} \\
L^2e^{2\beta_1\phi_1} &=&{1\over 1-\frac32{a^2\alpha^2}}\left[r^2
\left(1-\frac12{a^2\alpha^2}\right)-{a^2\Delta\over r^2} \right]\, ,
\label{twodilaton1}\\  e^{2\beta_2\phi_2}& =& r^2\, ,\label{twodilaton}\\
\Psi &=&{2 q \over  r}{a\over\sqrt{1-\frac12{a^2\alpha^2}\,}}\, ,
\end{eqnarray}
where $\beta_0, \bar\beta_0,\beta_1,\beta_2$ are arbitrary constants.
Even though this two-dimensional solution also presents interesting
properties, it will not be studied in detail here. For the
particular case with no charges and angular momentum see \cite{lemosCQG}.

\section{Conclusions}

We have presented the dimensional reduction to 3D of the rotating charged
toroidal-AdS black hole.
Dimensional reduction, through
the Killing azimuthal direction $\partial/\partial\varphi$,
 produced 3D black holes with an isotropic event horizon
(i.e., circularly symmetric), and the new charges were neatly found.

There are other interesting classes of black holes in 4D to which 
this procedure
could also be applied, namely the hyperbolic black holes
\cite{mann,peldan}, as well as the toroidal-AdS holes found in
\cite{klemm_mor_vanzo} which are not isometric to those of
\cite{lemos1,lz1}. Such an analysis can be done with the techniques 
presented here.

\section*{Acknowledgments}
This work is partially supported by FAPERGS (Funda\c c\~ao de Amparo \`a
Pesquisa do Estado do Rio Grande do Sul - Brazil).
One of us (VTZ) thanks Centro Multidisciplinar de Astrof\'\i sica at
Instituto Superior T\'ecnico (CENTRA-IST) for a grant and for hospitality.

\end{document}